\documentclass[onecolumn,nofootinbib]{revtex4}
\usepackage{graphicx}
\usepackage{amsmath}
\usepackage{amssymb}
\usepackage{float}
\usepackage[all]{xy}
\usepackage{amsfonts}
\usepackage{dcolumn}
\usepackage{bm}
\usepackage{xcolor}
\usepackage{caption}
\usepackage{hyperref}
\usepackage{multirow}
\usepackage[utf8x]{inputenc}
\usepackage{epstopdf}
\usepackage{capt-of}
\usepackage{dcolumn}   
\usepackage{bm}
\usepackage{dsfont}
\usepackage{relsize}
\usepackage{mathtools}

\begin{document}

\title{Energy Conditions in $f(P)$ Gravity }

\author{Snehasish Bhattacharjee \footnote{Email: snehasish.bhattacharjee.666@gmail.com}
  }\

\affiliation{Department of Physics, Indian Institute of Technology, Hyderabad 502285, India}
\date{\today}

\begin{abstract}
$f(P)$ gravity is a novel extension of ECG in which the Ricci scalar in the action is replaced by a function of the curvature invariant $P$ which represents the contractions of the Riemann tensor at the cubic order \cite{p}. The present work is concentrated on bounding some $f(P)$ gravity models using the concept of energy conditions where the functional forms of $f(P)$ are represented as \textbf{a)} $f(P) = \alpha \sqrt{P}$, and \textbf{b)} $f(P) = \alpha \exp (P)$, where $\alpha$ is the sole model parameter. Energy conditions are interesting linear relationships between pressure and density and have been extensively employed to derive interesting results in Einstein's gravity, and are also an excellent tool to impose constraints on any cosmological model. To place the bounds, we ensured that the energy density must remain positive, the pressure must remain negative, and the EoS parameter must attain a value close to $-1$ to make sure that the bounds respect the accelerated expansion of the Universe and are also in harmony with the latest observational data. We report that for both the models, suitable parameter spaces exist which satisfy the aforementioned conditions and therefore posit the $f(P)$ theory of gravity to be a promising modified theory of gravitation.

\end{abstract}

\maketitle
  
\section{Introduction}

Modified gravity theories are being continuously developed to explain the cosmic evolution from the primordial inflation just seconds after the big bang, to the formation of large-scale structures, and finally to the presently observed acceleration without citing dark matter and dark energy. Furthermore, in addition to being observationally harmonious \cite{p1,p2,p3,p4,p5}, modified gravity theories are quantizable gravitational theories and transpire naturally in the gravitational interactions of string theories \cite{p7, p6,p}, and encompasses Einstein's gravity as a distinct limit \cite{p8}.
$f(R)$ gravity was the first modified gravity introduced in literature where the Ricci scalar ($R$) is replaced by a suitable functional form of $R$, and is remarkable in explaining numerous cosmological observations \cite{2p4,2p5,2p6,2p7,2p8,2p9} (see \cite{fr} for a review on $f(R)$ gravity). Motivated by the successes of $f(R)$ gravity, several other gravitational theories emerged where the Ricci scalar is replaced by or with other curvature invariants such as the Gauss-Bonnet scalar, the Torsion scalar, and the Nonmetricity scalar (see \cite{rev} for a recent review on various modified gravity theories). \\
Recent years have seen a rapid development of a class of modified gravity theories employing cubic contractions of the Riemann tensor and is commonly termed Einsteinian cubic gravity (ECG) \cite{p31}. Ref \cite{p} reported a novel extension of ECG in which the Ricci scalar in the action is replaced by a function of the curvature invariant $P$ which represents the contractions of the Riemann tensor at the cubic order. This class of modified gravity theory have been very successful in several cosmological domains such as inflation \cite{p35}, spherically symmetric black hole solutions \cite{p32,p33,p34}, and late-time acceleration \cite{p,2p}. \\
Energy conditions are interesting linear relationships between pressure and density and have been extensively employed to derive interesting results in Einstein's gravity. For instance, the Hawking-Penrose singularity theorems employ the strong energy condition (SEC) \cite{ft33}, whose violation warrants an accelerated expansion of the cosmos. Another example is the invoking of the null energy condition (NEC) to prove the second law of black hole thermodynamics \cite{ft34,ft35}.  \\
Energy conditions are an excellent tool to impose constraints on any cosmological model \cite{ft36}. Energy conditions have been used to derive constraints on several modified gravity theories \cite{ft45,ft46,ft47,ft48,ft49,ft50,ft51,ft52,bhatta,jamil,garcia}. They are also employed to bound several cosmological observables such as the lookback time, curvature parameter, distance modulus, and the deceleration parameter \cite{ft37,ft38,ft39,ft40,ft41,ft42,ft43,ft44}.\\
In this work, we shall derive the expressions for the energy conditions in $f(P)$ gravity to impose strict bounds on this class of modified gravity by employing the observational constraints on the Hubble, the deceleration, and the jerk parameters. The manuscript is organized as follows: In Section \ref{sec2}, we provide an overview of $f(P)$ gravity and obtain the expressions of the energy density, pressure, and the EoS parameter. In Section \ref{sec3}, we summarize the concept of energy conditions and obtain the expressions for the energy conditions in $f(P)$ gravity. In Section \ref{sec4} we impose bounds on some $f(P)$ gravity models using the concept of energy conditions and in Section \ref{sec5} we present the results and conclusions. Throughout the work we shall work with natural units and use $H=0.692$, $q=-0.545$, and $j=0.776$ \cite{cap/2019}.

\section{$f(P)$ Gravity: An overview}\label{sec2}

The nontopological cubic term $P$ is defined as \cite{p}
\begin{multline}
P = \beta_{1} R_{i}{}^{\rho}{}_{j}{}^{\sigma}R_{\rho}{}^{\gamma}{}_{\sigma}{}^{\delta}R_{\gamma}{}^{i}{}_{\delta}{}^{j}+\beta_{2} R R_{ij\rho \sigma}R^{ij\rho \sigma}+\beta_{3} R^{j}_{i} R^{\rho}_{j}     R^{i}_{\rho}  +\beta_{4}R^{\rho \sigma}_{ij}R^{\gamma \delta}_{\rho \sigma} R^{ij}_{\gamma \delta}\\+\beta_{5} R^{i\rho}R^{j\sigma}R_{ij \rho \sigma}  +\beta_{6} R ^{\sigma \gamma}R_{ij\rho \sigma}R^{ij\rho \sigma}{}_{\gamma} + \beta_{7} R R^{ij} R_{ij} +\beta_{8} R^{3},
\end{multline}

where $\beta_{i}$ are free parameters. Now, to ensure that $f(P)$ theories have a spectrum similar to that of Einstein's gravity, the parameters $\beta_{7}$, and $\beta_{8}$ need to take the following forms \cite{p31} 

\begin{equation}
\beta_{7} = \frac{1}{12}\left[3 \beta_{1} -48 \beta_{2} -9 \beta_{3} -24 \beta_{4} - 5 \beta_{5} - 16 \beta_{6}  \right], 
\end{equation}
\begin{equation}
\beta_{8} = \frac{1}{72}\left[-6 \beta_{1}  + 64 \beta_{2}  + 9  \beta_{3} + 36\beta_{4} +3 \beta_{5} + 22 \beta _{6}\right]. 
\end{equation}

The action in $f(P)$ gravity reads \cite{p}

\begin{equation}\label{action}
\mathcal{S} - \frac{1}{2 \kappa}\int \sqrt{-g}\left( R + f(P)\right) d^{4}x, 
\end{equation}
where $g$ represents the metric determinant, $R$ the Ricci scalar, $\kappa = 8 \pi G$ the Newton's gravitational constant, and $f(P)$ is a suitable functional choice of $P$. Varying the action (Eq. \ref{action}) yields the following field equation 
\begin{equation}\label{field}
G_{ij} = \kappa (\bar{H}_{ij} + T_{ij}),
\end{equation}
where $T_{ij}$ is the energy-momentum tensor and is defined as 
\begin{equation}
T_{ij} = - \left(\frac{2}{\sqrt{-g}} \right) \left[\frac{\delta(\mathcal{L}_{m}\sqrt{-g})}{\delta g^{ij}} \right], 
\end{equation}
and 
\begin{equation}
\bar{H}_{ij}  = R^{\epsilon\varepsilon\zeta}(_{i}\bar{K}_{j})_{\zeta\epsilon\varepsilon} + g_{ij}f(P)\nabla^{\epsilon}\nabla^{\varepsilon} \bar{K}_{\epsilon(ij)\varepsilon}.
\end{equation}
The tensor $\bar{K}_{\epsilon\varepsilon ij}$ is defined as $K_{\epsilon\varepsilon ij} f^{'}(P)$, where $f^{'}(P) = df/dP$, and $K_{\epsilon\varepsilon ij}$ can be expressed as \cite{p} 
\begin{equation}
K_{\epsilon\varepsilon ij} = 12 \left[ 6 R_{\epsilon}{}^{\zeta}(_{i}{}^{\sigma}R_{\nu})_{\sigma \varepsilon \zeta} + \frac{1}{2} R_{\epsilon \varepsilon}^{\zeta \sigma} R_{ij \zeta \sigma} + 2g_{\varepsilon} (_{i}{}R_{\nu})_{\sigma \epsilon \zeta} R ^{\zeta \sigma} - 4R_{\zeta}(_{i}g_{j})(_{\epsilon}R_{\varepsilon})^{\zeta} -2 g_{\epsilon}(_{i}R_{j})_{\sigma \varepsilon \zeta} R^{\zeta \sigma}  - 2 R _{\epsilon} (_{\epsilon}R_{\varepsilon})_{j}\right]. 
\end{equation}
For a FLRW background, the field equations (Eq. \ref{field}) take the following forms \cite{p}

\begin{equation}
3H^{2} = \kappa \rho_{eff},
\end{equation} 
\begin{equation}
2 \dot{H} + 3H^{2} = -\kappa p_{eff},
\end{equation}
where, $\rho_{eff} = \rho_{m} + \rho_{f(P)}$, and $p_{eff} =p_{m} + p_{f(P)}$. Furthermore, $\rho_{f(P)}$, and $p_{f(P)}$ are defined respectively as \cite{p}
\begin{equation}\label{rho}
\rho_{f(P)} = - f(P) + 18 \beta H^{4} (\dot{H} - H \frac{\partial}{\partial t} + H^{2}) f^{'}(P),
\end{equation}
\begin{equation}\label{p}
p_{f(P)} = f(P) + 6  \beta H^{3} \left[ 2(H^{2} + 2\dot{H})\frac{\partial}{\partial t} -5H\dot{H} + H\frac{\partial^{2}}{\partial t^{2}} - 3 H^{3}  \right] f^{'}(P).
\end{equation}

The EoS parameter ($\omega_{f(P)}=p_{f(P)}/\rho_{f(P)} $) therefore reads \begin{equation}\label{om}
\omega_{f(P)} = \left[\frac{f(P) + 6  \beta H^{3} \left\lbrace  2(H^{2} + 2\dot{H})\frac{\partial}{\partial t} -5H\dot{H} + H\frac{\partial^{2}}{\partial t^{2}} - 3 H^{3}  \right\rbrace  f^{'}(P)}{- f(P) + 18  \beta H^{4} (\dot{H} - H \frac{\partial}{\partial t} + H^{2}) f^{'}(P)} \right]. 
\end{equation}

Additionally, for a FRW metric, $P = 6  \beta H^{4} (2 H^{2} + 3 \dot{H})$, and $ \beta $ is defined as 
\begin{equation}
 \beta  = - \beta_{1}+ 8 \beta_{2} + 2 \beta_{3} + 4 \beta_{4}.
\end{equation}

\section{Energy Conditions in $f(P)$ Gravity}\label{sec3}

The energy conditions transpire from the Raychaudhuri equation for the cosmic expansion defined as 
\begin{equation}\label{ray}
\frac{d \theta}{d\tau} = \omega_{ij}\omega^{ij}-\left[ R_{ij}u^{i}u^{j} + \frac{1}{3} \theta^{2} + \sigma_{ij}\sigma^{ij}\right], 
\end{equation}

where $\theta$, $\omega_{ij}$, and $\sigma^{ij}$ are the expansion, rotation and shear coupled to the congruence described by the vector $u^{i}$, and $R_{ij}$ denote the Ricci tensor. \\

Additionally, the evolutionary expansion for the congruence of null geodesics expressed through a null vector $\kappa^{i}$ reads 
\begin{equation}\label{ray2}
\frac{d \theta}{d\tau} = \omega_{ij}\omega^{ij}-\left[ R_{ij}\kappa^{i}\kappa^{j} + \frac{1}{2} \theta^{2} + \sigma_{ij}\sigma^{ij}\right], 
\end{equation}
where $\theta$, $\omega_{ij}$, and $\sigma^{ij}$ are now coupled to the null geodesics.

One may note that the Eqs. \ref{ray} and \ref{ray2} are solely geometric in nature and are identical for all gravitational theories.\\
It is also evident that $\sigma_{ij}\sigma^{ij}\geq 0$ since the shear is a spatial tensor and therefore, from Eqs. \ref{ray} and \ref{ray2} it is clear that for any hypersurface orthogonal congruences described by $\omega_{ij} = 0$, the necessary criteria required to establish  gravity as an attractive force read $R^{ij}u^{i}u^{j} \geq 0$ and $R^{ij}\kappa^{i}\kappa^{j} \geq 0$ \cite{ft48}. These inequalities further corroborate the fact that the geodesic congruences pivot within some finite values of the parameters on the geodesics \cite{ft48}. We shall now try to derive the energy conditions in $f(P)$ gravity using the modified gravitational equations \ref{field}. The energy conditions are self-governing and are entirely geometrical in nature \cite{jamil,ft33} and can be expressed as follows \cite{garcia,ft41}:

\begin{equation}\label{nec1}
\textbf{\textit{Null Energy Condition} (NEC):} \hspace{0.1in} \rho_{eff} + p_{eff} \geq 0,
\end{equation}
\begin{equation}\label{sec1}
\textbf{\textit{Strong Energy Condition} (SEC):} \hspace{0.1in} \rho_{eff} + 3p_{eff} \geq 0, \hspace{0.1in} \text{and} \hspace{0.1in} \rho_{eff} + p_{eff} \geq 0,
\end{equation}
\begin{equation}\label{wec1}
\textbf{\textit{Weak Energy Condition} (WEC):} \hspace{0.1in}\rho_{eff} \geq 0, \hspace{0.1in} \text{and} \hspace{0.1in} \rho_{eff} + p_{eff} \geq 0.
\end{equation}

Now, in order to proceed, it is useful to re-express the deceleration ($q$), and jerk parameters ($j$) in terms of the Hubble parameter  as follows:
\begin{equation}\label{q}
q = -1\left[\frac{1}{H^{2}} \frac{\dot{a}}{a} \right], 
\end{equation}
and
\begin{equation}\label{j}
j = \left[\frac{1}{H^{3}} \frac{\ddot{a}}{a} \right]. 
\end{equation}

Using Eq. \ref{q} and Eq. \ref{j}, we can now write the time derivatives of $H$ in terms of $q$ and $j$ as follows:
\begin{equation}\label{dh}
\dot{H} = - H^{2}(q+1),
\end{equation}
and
\begin{equation}\label{ddh}
\ddot{H} = H^{4} (j + 2 + 3q).
\end{equation}

It may be noted that the matter sector (i.e, $\rho_{m}$ and $p_{m}$) do not violate the energy conditions and that the violation of energy conditions is solely regulated by $\rho_{fp}$ and $p_{fp}$. Therefore, by substituting Eqs. \ref{dh}, and \ref{ddh} in Eqs. \ref{rho}, and \ref{p}, we define the energy conditions in $f(P)$ theories of gravity as follows: 

\begin{multline}\label{nec}
\textbf{NEC:} \hspace{0.1in} - f(P) - 18  \beta H^{4} ( H^{2}(q+1) - H \frac{\partial}{\partial t} + H^{2}) f^{'}(P)  \\ + f(P) + 6  \beta H^{3} \left[ 2(H^{2} - 2\left\lbrace  H^{2}(q+1))\right\rbrace \frac{\partial}{\partial t} +5H( H^{2}(q+1)) + H\frac{\partial^{2}}{\partial t^{2}} - 3 H^{3}  \right] f^{'}(P) \geq 0
\end{multline}  

\begin{multline}\label{sec}
\textbf{SEC:} \hspace{0.1in} - f(P) - 18  \beta H^{4} ( H^{2}(q+1) - H \frac{\partial}{\partial t} + H^{2}) f^{'}(P)  \\ + 3\left[ f(P) + 6  \beta H^{3} \left[ 2(H^{2} - 2\left\lbrace  H^{2}(q+1))\right\rbrace \frac{\partial}{\partial t} +5H( H^{2}(q+1)) + H\frac{\partial^{2}}{\partial t^{2}} - 3 H^{3}  \right] f^{'}(P)\right]  \geq 0, \\ \text{and} \\
\hspace{0.1in} - f(P) - 18  \beta H^{4} ( H^{2}(q+1) - H \frac{\partial}{\partial t} + H^{2}) f^{'}(P)  \\ + f(P) + 6  \beta H^{3} \left[ 2(H^{2} - 2\left\lbrace  H^{2}(q+1))\right\rbrace \frac{\partial}{\partial t} +5H( H^{2}(q+1)) + H\frac{\partial^{2}}{\partial t^{2}} - 3 H^{3}  \right] f^{'}(P) \geq 0
\end{multline}  

\begin{multline}\label{wec}
\textbf{WEC:} \hspace{0.1in} - f(P) - 18  \beta H^{4} ( H^{2}(q+1) - H \frac{\partial}{\partial t} + H^{2}) f^{'}(P)  \geq 0, \\ \text{and} \\
\hspace{0.1in} - f(P) - 18  \beta H^{4} ( H^{2}(q+1) - H \frac{\partial}{\partial t} + H^{2}) f^{'}(P)  \\ + f(P) + 6  \beta H^{3} \left[ 2(H^{2} - 2\left\lbrace  H^{2}(q+1))\right\rbrace \frac{\partial}{\partial t} +5H( H^{2}(q+1)) + H\frac{\partial^{2}}{\partial t^{2}} - 3 H^{3}  \right] f^{'}(P) \geq 0.
\end{multline}  

\section{EC bounds on $f(P)$ theories}\label{sec4}
We shall now use Eqs. \ref{nec}, \ref{sec} and \ref{wec} to bound some suitable $f(P)$ gravity models. To place the bounds, we shall ensure that the energy density (Eq. \ref{rho}) must remain positive and the pressure (Eq. \ref{p}) must be negative to assure that the constraints on $f(P)$ gravity models are compatible with the current accelerated expansion of the Universe. Next, we shall investigate to find corners in parameter spaces for a chosen model which permits the EoS parameter (Eq. \ref{om}) to attain a value close to $-1$ to ensure that the bounds are in harmony with the current observational data. We note that for an accelerating Universe, the SEC must violate and therefore offers a consistency check to the constraints obtained for the models.\\
In the subsequent subsections, we shall try to place bounds for two specific $f(P)$ gravity models where the functional forms of $f(P)$ are represented as \textbf{a)} $f(P) = \alpha \sqrt{P}$, and \textbf{b)} $f(P) = \alpha \exp (P)$, where $\alpha$ is the sole model parameter for these models.

\subsection{$f(P) = \alpha \sqrt{P}$} 

For the first case, let us assume the functional form of $f(P)$ is given by 
\begin{equation}
f(P) = \alpha \sqrt{P},
\end{equation}
where $\alpha$ is the model parameter. \\

The expressions for the NEC (Eq. \ref{nec}), SEC (Eq. \ref{sec}), and WEC (Eq. \ref{wec}) read respectively as 
\begin{multline}
\textbf{NEC:} \hspace{0.1in} -\left[ \frac{\sqrt{\frac{3}{2}} \alpha  \beta  H^6 (6 q+7)}{\sqrt{\beta  \left(-H^6\right) (3 q+1)}}\right] \\+ \frac{\beta  H^5 \left[ 60 H (q+1) \sqrt{\beta  \left(-H^6\right) (3 q+1)}+\sqrt{6} \alpha  \left\lbrace 12 (j+3 q+2)-H (12 q (q+2)+11)\right\rbrace \right] }{2 \sqrt{\beta  \left(-H^6\right) (3 q+1)}} \geq 0,
\end{multline}
\begin{multline}
\textbf{SEC:} \hspace{0.1in} -\left[ \frac{\sqrt{\frac{3}{2}} \alpha  \beta  H^6 (6 q+7)}{\sqrt{\beta  \left(-H^6\right) (3 q+1)}}\right] \\+ 3\left[ \frac{\beta  H^5 \left[ 60 H (q+1) \sqrt{\beta  \left(-H^6\right) (3 q+1)}+\sqrt{6} \alpha  \left\lbrace 12 (j+3 q+2)-H (12 q (q+2)+11)\right\rbrace \right] }{2 \sqrt{\beta  \left(-H^6\right) (3 q+1)}}\right]  \geq 0, \\
\text{and} \\
 -\left[ \frac{\sqrt{\frac{3}{2}} \alpha  \beta  H^6 (6 q+7)}{\sqrt{\beta  \left(-H^6\right) (3 q+1)}}\right] \\+ \frac{\beta  H^5 \left[ 60 H (q+1) \sqrt{\beta  \left(-H^6\right) (3 q+1)}+\sqrt{6} \alpha  \left\lbrace 12 (j+3 q+2)-H (12 q (q+2)+11)\right\rbrace \right] }{2 \sqrt{\beta  \left(-H^6\right) (3 q+1)}} \geq 0,
\end{multline}
\begin{multline}
\textbf{WEC:} \hspace{0.1in} -\left[ \frac{\sqrt{\frac{3}{2}} \alpha  \beta  H^6 (6 q+7)}{\sqrt{\beta  \left(-H^6\right) (3 q+1)}}\right] \geq 0  \\
 \text{and} \\
  -\left[ \frac{\sqrt{\frac{3}{2}} \alpha  \beta  H^6 (6 q+7)}{\sqrt{\beta  \left(-H^6\right) (3 q+1)}}\right] \\+ 3\frac{\beta  H^5 \left[ 60 H (q+1) \sqrt{\beta  \left(-H^6\right) (3 q+1)}+\sqrt{6} \alpha  \left\lbrace 12 (j+3 q+2)-H (12 q (q+2)+11)\right\rbrace \right] }{2 \sqrt{\beta  \left(-H^6\right) (3 q+1)}} \geq 0.
\end{multline}

The expressions for the energy density $(\rho_{f(P)})$ and pressure ($p_{f(P)}$) for this model read respectively as 
\begin{equation}\label{rho1}
\rho_{f(P)} = -\left[ \frac{\sqrt{\frac{3}{2}} \alpha  \beta  H^6 (6 q+7)}{\sqrt{\beta  \left(-H^6\right) (3 q+1)}}\right] ,
\end{equation}

\begin{equation}\label{p1}
p_{f(P)}= \frac{\beta  H^5 \left[ 60 H (q+1) \sqrt{\beta  \left(-H^6\right) (3 q+1)}+\sqrt{6} \alpha  \left\lbrace 12 (j+3 q+2)-H (12 q (q+2)+11)\right\rbrace \right] }{2 \sqrt{\beta  \left(-H^6\right) (3 q+1)}}.
\end{equation}

The EoS parameter ($\omega_{f(P)}$) reads

\begin{equation}\label{om1}
\omega_{f(P)}  =  \frac{\sqrt{6} \alpha  (H (12 q (q+2)+11)-12 (j+3 q+2))-60 H (q+1) \sqrt{\beta  \left(-H^6\right) (3 q+1)}}{\sqrt{6} \alpha  H (6 q+7)}.
\end{equation}
Substituting the values of $H$, $q$ and $j$ in Eqs. \ref{rho1}, \ref{p1}, and \ref{om1}, we obtain 

\begin{equation}\label{rho2}
\rho_{f(P)} \simeq-1.9 \alpha  \sqrt{\beta }
\end{equation}
\begin{equation}\label{p2}
p_{f(P)}\simeq 9.2 \alpha  \sqrt{\beta }+1.5 \beta
\end{equation} 
and 
\begin{equation}\label{om2}
\omega_{f(P)} \simeq  -\frac{0.8 \sqrt{\beta }}{\alpha }-5.0.
\end{equation}

From Eqs. \ref{rho2}, \ref{p2}, and \ref{om2} it is clear that:
\begin{itemize}
\item For $\rho_{f(P)}\geq 0$: $\alpha \leq 0$, and $\beta \geq 0$.
\item For $p_{f(P)} \leq 0$: $\alpha \leq 0$, $\beta \geq 0$ and $\frac{\alpha}{\sqrt{\beta}} \lesssim -0.15$.
\item For $\omega_{f(P)} \simeq -1$:  $\alpha \leq 0$, $\beta \geq 0$ and $\frac{\alpha}{\sqrt{\beta}} \simeq -0.2$.
\end{itemize} 

\begin{figure}[H]
\centering
  \includegraphics[width=7.5 cm]{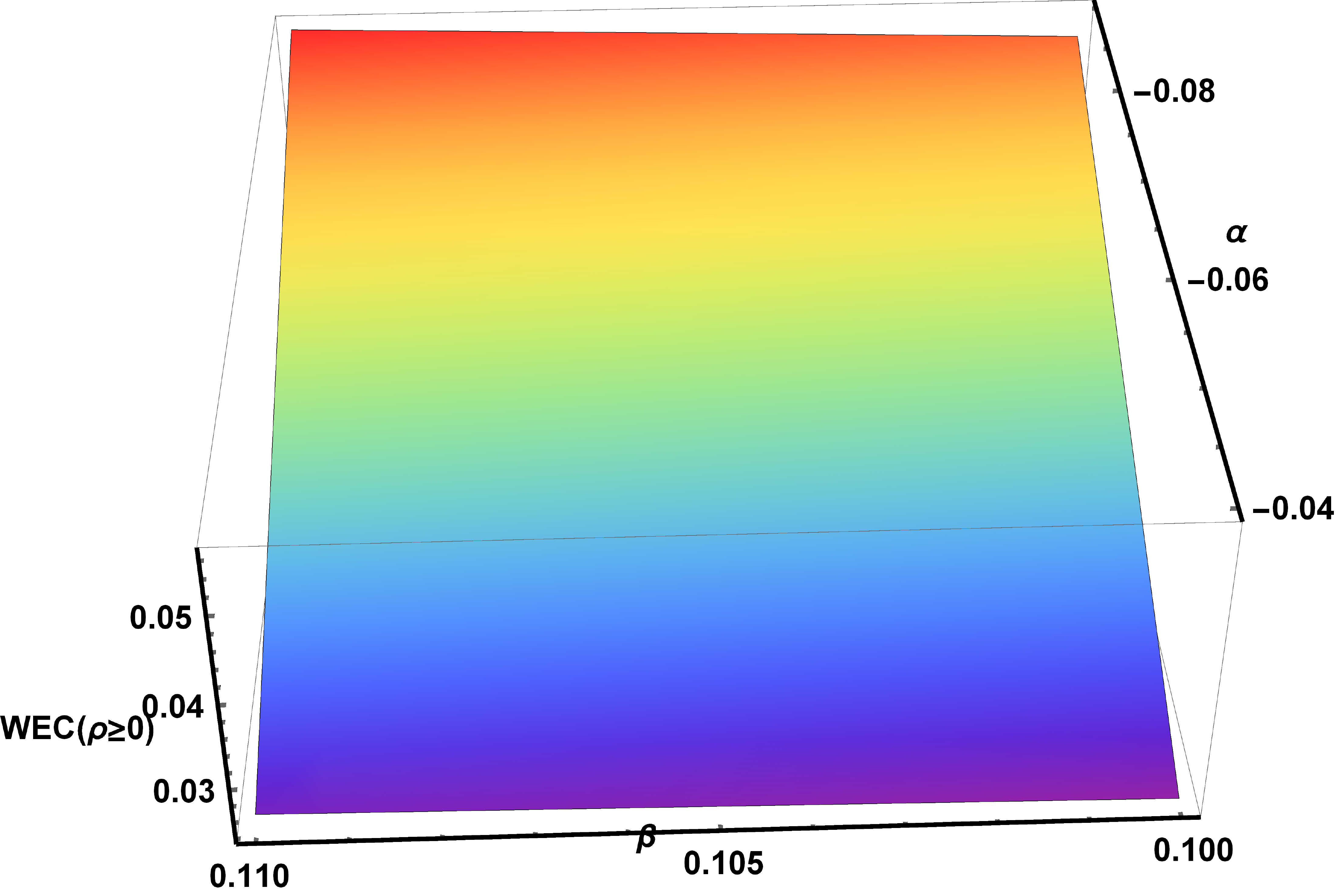}  
  \includegraphics[width=7.5 cm]{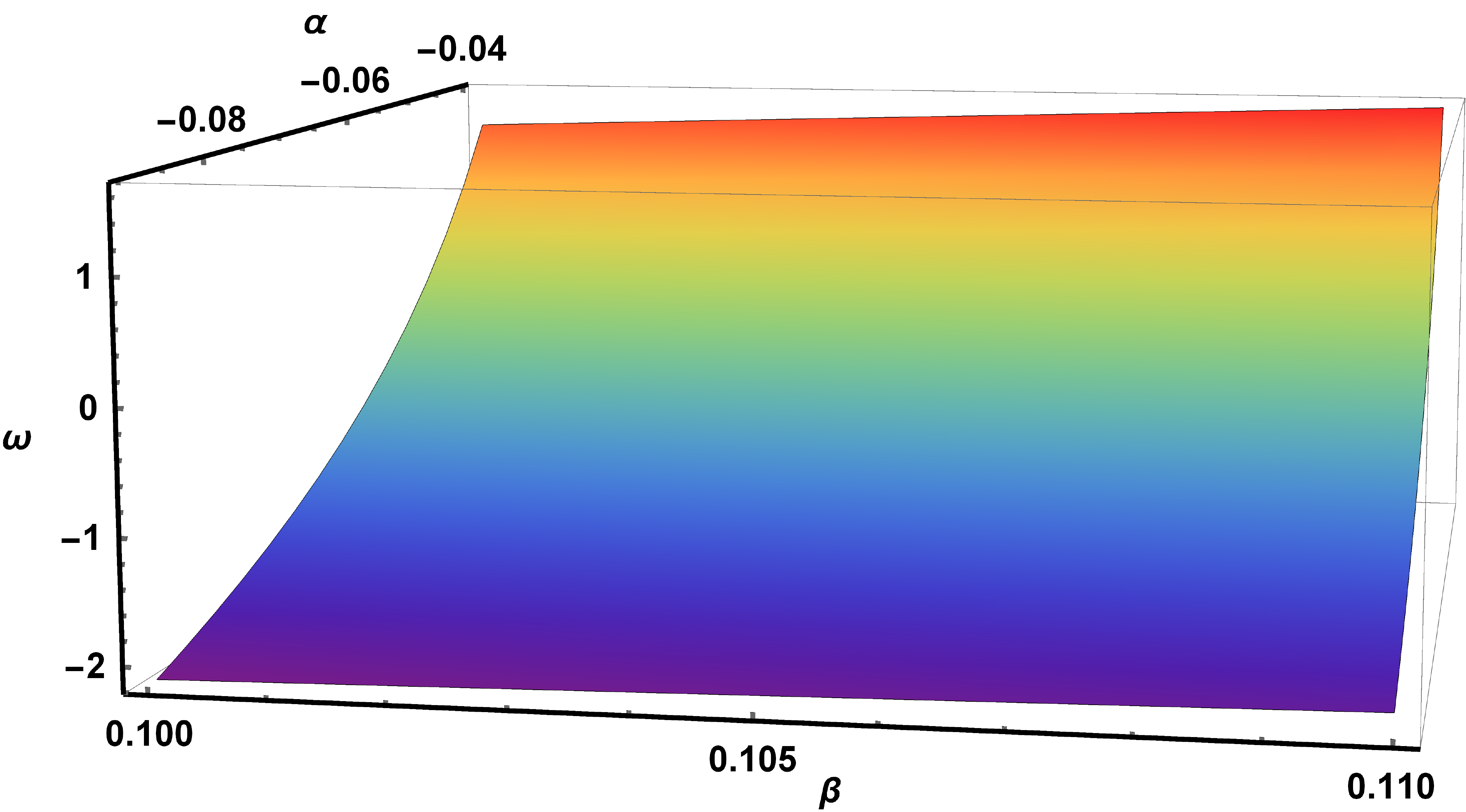}
  \includegraphics[width=7.5 cm]{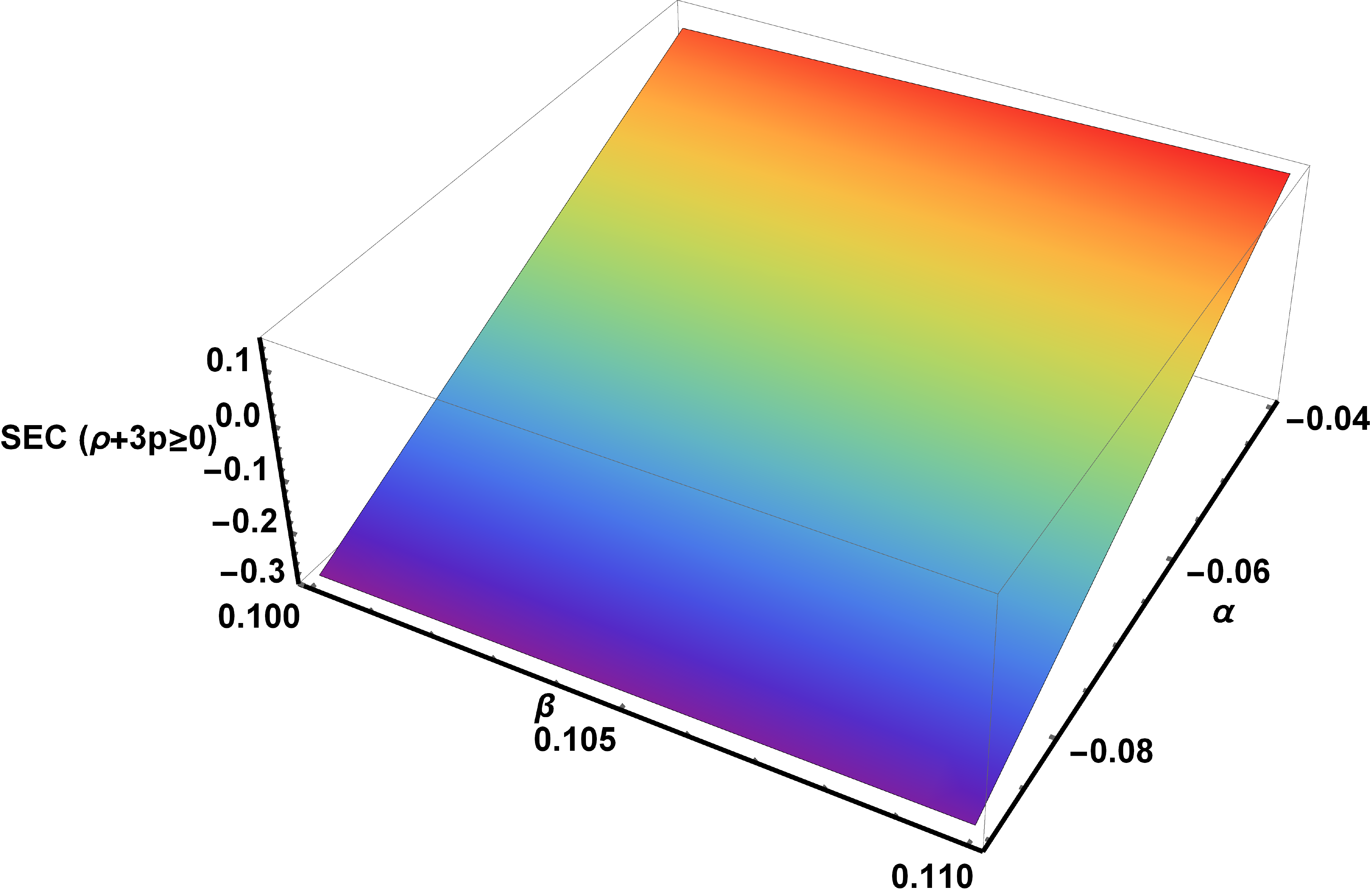}
  \includegraphics[width=7.5 cm]{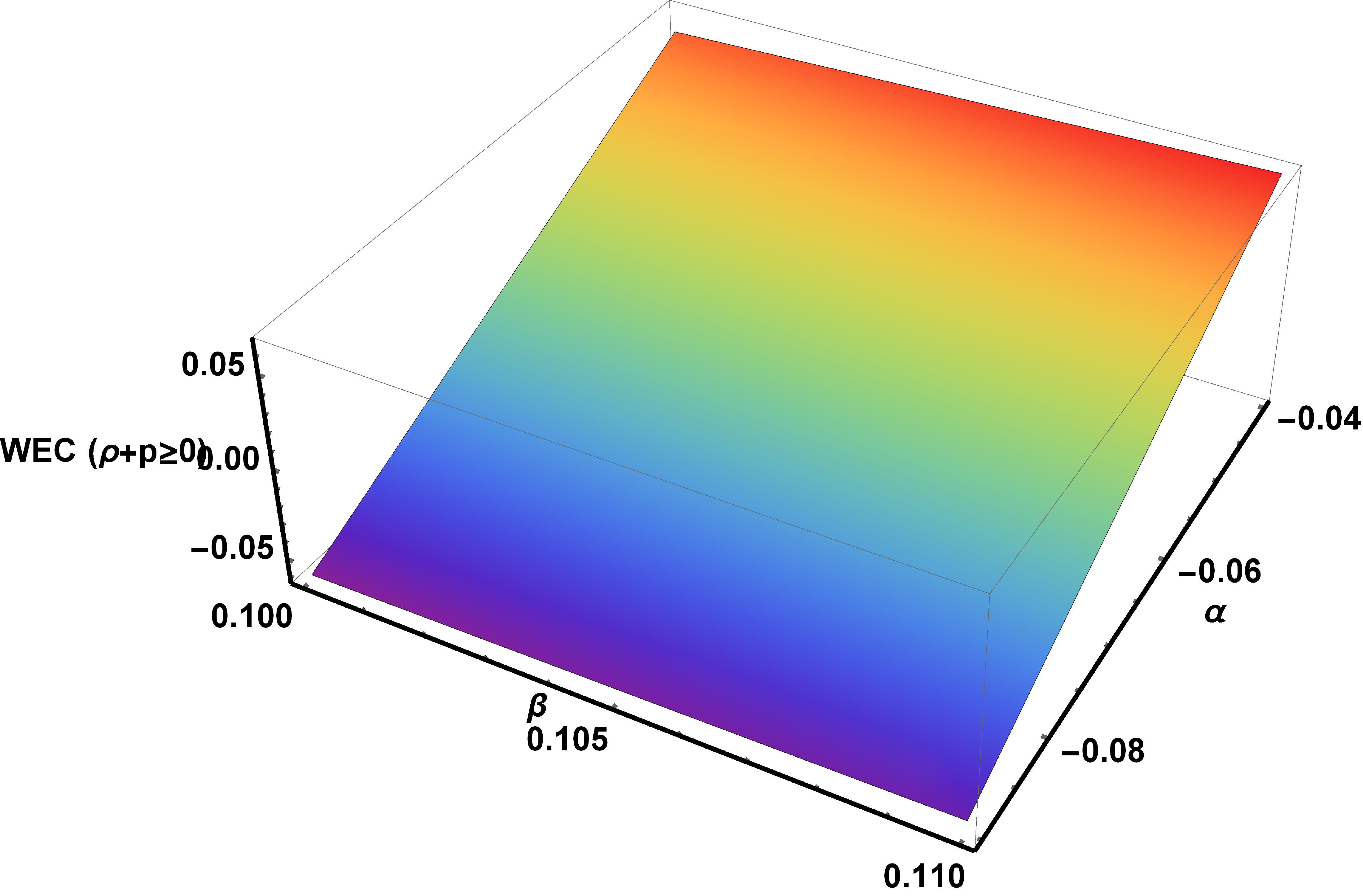}
\caption{Top left panel shows the profile of energy density $\rho_{f(P)}$, top right panel shows the profile of the EoS parameter $\omega_{f(P)}$, bottom left panel shows the profile of the Strong Energy Condition and the bottom right panel shows the profile of the Weak Energy Condition for the first ($f(P) = \alpha \sqrt{P}$) model. The figures are drawn for $ \beta \left\lbrace 0.1 , 0.11 \right\rbrace$, and $ \alpha  \left\lbrace -0.04, -0.09 \right\rbrace$. }
\label{fig1}
\end{figure}

\subsection{$f(P) = \alpha \exp (P)$}

Let us now set the functional form of $f(P)$ to the following 
\begin{equation}
f(P) = \alpha \exp (P),
\end{equation}
with $\alpha$ being the model parameter. \\

Similar to the previous case, we shall express the NEC (Eq. \ref{nec}), SEC (Eq. \ref{sec}), and WEC (Eq. \ref{wec}) for this model respectively as 

\begin{multline}
\textbf{NEC:} \hspace{0.1in} \left[ \alpha  e^{-6 \beta  H^6 (3 q+1)} \left\lbrace 18 \beta  H^6 \left(-36 \beta  H^6 (q+1) (3 q+1)-q\right)-1\right\rbrace \right] \\ + \left[ \splitfrac{ \left[ 30 \beta  H^6 (q+1)+\alpha  e^{-6 \beta  H^6 (3 q+1)}\right] }{\times\left[  \left(18 \beta  H^6 \left( 12 \beta  H^5 (3 q+1) \left(36 \beta  H^6 (3 q+1) (j+3 q+2)-2 H (q+1) (2 q+1)-5 (j+3 q+2)\right)-1\right)+1\right)\right] } \right] \geq 0,
\end{multline}
\begin{multline}
\textbf{SEC:} \hspace{0.1in} \left[ \alpha  e^{-6 \beta  H^6 (3 q+1)} \left\lbrace 18 \beta  H^6 \left(-36 \beta  H^6 (q+1) (3 q+1)-q\right)-1\right\rbrace \right] \\+ 3 \left[ \splitfrac{ \left[ 30 \beta  H^6 (q+1)+\alpha  e^{-6 \beta  H^6 (3 q+1)}\right] }{\times\left[  \left(18 \beta  H^6 \left( 12 \beta  H^5 (3 q+1) \left(36 \beta  H^6 (3 q+1) (j+3 q+2)-2 H (q+1) (2 q+1)-5 (j+3 q+2)\right)-1\right)+1\right)\right] } \right]  \geq 0, \\
\text{and} \\
\left[ \alpha  e^{-6 \beta  H^6 (3 q+1)} \left\lbrace 18 \beta  H^6 \left(-36 \beta  H^6 (q+1) (3 q+1)-q\right)-1\right\rbrace \right]\\ + \left[ \splitfrac{ \left[ 30 \beta  H^6 (q+1)+\alpha  e^{-6 \beta  H^6 (3 q+1)}\right] }{\times\left[  \left(18 \beta  H^6 \left( 12 \beta  H^5 (3 q+1) \left(36 \beta  H^6 (3 q+1) (j+3 q+2)-2 H (q+1) (2 q+1)-5 (j+3 q+2)\right)-1\right)+1\right)\right] }\right]  \geq 0,
\end{multline}
\begin{multline}
\textbf{WEC:} \hspace{0.1in} \left[ \alpha  e^{-6 \beta  H^6 (3 q+1)} \left\lbrace 18 \beta  H^6 \left(-36 \beta  H^6 (q+1) (3 q+1)-q\right)-1\right\rbrace \right] \geq 0  \\
 \text{and} \\
 \left[ \alpha  e^{-6 \beta  H^6 (3 q+1)} \left\lbrace 18 \beta  H^6 \left(-36 \beta  H^6 (q+1) (3 q+1)-q\right)-1\right\rbrace \right] \\ + \left[ \splitfrac{ \left[ 30 \beta  H^6 (q+1)+\alpha  e^{-6 \beta  H^6 (3 q+1)}\right] }{\times\left[  \left(18 \beta  H^6 \left( 12 \beta  H^5 (3 q+1) \left(36 \beta  H^6 (3 q+1) (j+3 q+2)-2 H (q+1) (2 q+1)-5 (j+3 q+2)\right)-1\right)+1\right)\right] } \right] \geq 0.
\end{multline}

The expressions for the energy density $(\rho_{f(P)})$ and pressure ($p_{f(P)}$) for the second model read respectively as 
\begin{equation}\label{rho3}
\rho_{f(P)} = \left[ \alpha  e^{-6 \beta  H^6 (3 q+1)} \left\lbrace 18 \beta  H^6 \left(-36 \beta  H^6 (q+1) (3 q+1)-q\right)-1\right\rbrace \right]  ,
\end{equation}

\begin{equation}\label{p3}
p_{f(P)}= \splitfrac{ \left[ 30 \beta  H^6 (q+1)+\alpha  e^{-6 \beta  H^6 (3 q+1)}\right] }{\times\left[  \left(18 \beta  H^6 \left( 12 \beta  H^5 (3 q+1) \left(36 \beta  H^6 (3 q+1) (j+3 q+2)-2 H (q+1) (2 q+1)-5 (j+3 q+2)\right)-1\right)+1\right)\right] }.
\end{equation}

The EoS parameter ($\omega_{f(P)}$) for this model reads

\begin{equation}\label{om3}
\omega_{f(P)}  =  \frac{\splitfrac{ \left[ 30 \beta  H^6 (q+1)+\alpha  e^{-6 \beta  H^6 (3 q+1)}\right] }{\times\left[  \left(18 \beta  H^6 \left( 12 \beta  H^5 (3 q+1) \left(36 \beta  H^6 (3 q+1) (j+3 q+2)-2 H (q+1) (2 q+1)-5 (j+3 q+2)\right)-1\right)+1\right)\right] }}{\alpha  e^{-6 \beta  H^6 (3 q+1)} \left(18 \beta  H^6 \left(-36 \beta  H^6 (q+1) (3 q+1)-q\right)-1\right)}.
\end{equation}
We now substitute the respective values of $H$, $q$ and $j$ into \ref{rho3}, \ref{p3}, and \ref{om3} to obtain 

\begin{equation}\label{rho4}
\rho_{f(P)} \simeq \alpha  e^{0.42 \beta } (\beta  (2.3 \beta +1.1)-1)
\end{equation}
\begin{equation}\label{p4}
p_{f(P)}\simeq \alpha  e^{0.3 \beta } (\beta  (\beta  (6.9 \beta +13.5)-2.0)+1.)+1.5 \beta
\end{equation} 
and 
\begin{equation}\label{om4}
\omega_{f(P)} \simeq  -\frac{\alpha  (\beta  (\beta  (6.9 \beta +13.5)-2.0)+1.)+1.5 e^{-0.4 \beta } \beta }{\alpha  (\beta  (-2.3 \beta -1.1)+1)}.
\end{equation}

Given the complexity of Eqs. \ref{rho4}, \ref{p4}, and \ref{om4}, it is therefore difficult to constrain $\beta$ and 
$\alpha$. Let us therefore proceed by assuming first $\beta = +1$, and then $\beta = -1$ and check for which of these two cases, we get $\rho_{f(P)} \geq 0$, and $p_{f(P)} \leq 0$.\\\\
\textbf{Case 1:} $\beta = + 1$ gives
\begin{equation}\label{rho5}
\rho_{f(P)} \simeq 3.4 \alpha,
\end{equation}
and
\begin{equation}\label{p5}
p_{f(P)} \simeq 29.3 \alpha +1.5.
\end{equation}

From Eq. \ref{rho5}, and \ref{p5} we get the following constraints for $\alpha$:
\begin{itemize}
\item For $\rho_{f(P)}\geq 0: \alpha \geq 0$,
\item For $p_{f(P)} \leq 0$: $\alpha \leq 0$.
\end{itemize}
Since to obtain  $\rho_{f(P)}\geq 0$, and $p_{f(P)} \leq 0$, $\alpha$ needs to be positive and negative at the same time which is unphysical, it is therefore clear that positive values for $\beta$ are not permissible for this model. \\\\
\textbf{Case 2:} $\beta = -1$ gives
\begin{equation}\label{rho6}
\rho_{f(P)} \simeq 0.2 \alpha,
\end{equation}
and
\begin{equation}\label{p6}
p_{f(P)} \simeq 6.2 \alpha -1.5.
\end{equation}
Eqs. \ref{rho6}, and \ref{p6} result in the following constraints for $\alpha$:
\begin{itemize}
\item For $\rho_{f(P)}\geq 0: \alpha \geq 0$,
\item For $p_{f(P)} \leq 0$: $ 0 \leq \alpha \lesssim 0.24$.
\end{itemize}
Therefore, it is evident that there exist suitable values for $\alpha$ which permit an accelerating Universe for this $f(P)$ gravity model given that $\beta$ remain negative. Finally $\omega_{f(P)}$ for this model reads
\begin{equation}\label{om6}
\omega_{f(P)} \simeq 53.3 -\frac{12.5}{\alpha },
\end{equation}
where we report that for Eq. \ref{om6} $\simeq -1$, $\alpha \simeq 0.23$.

\begin{figure}[H]
\centering
  \includegraphics[width=7.5 cm]{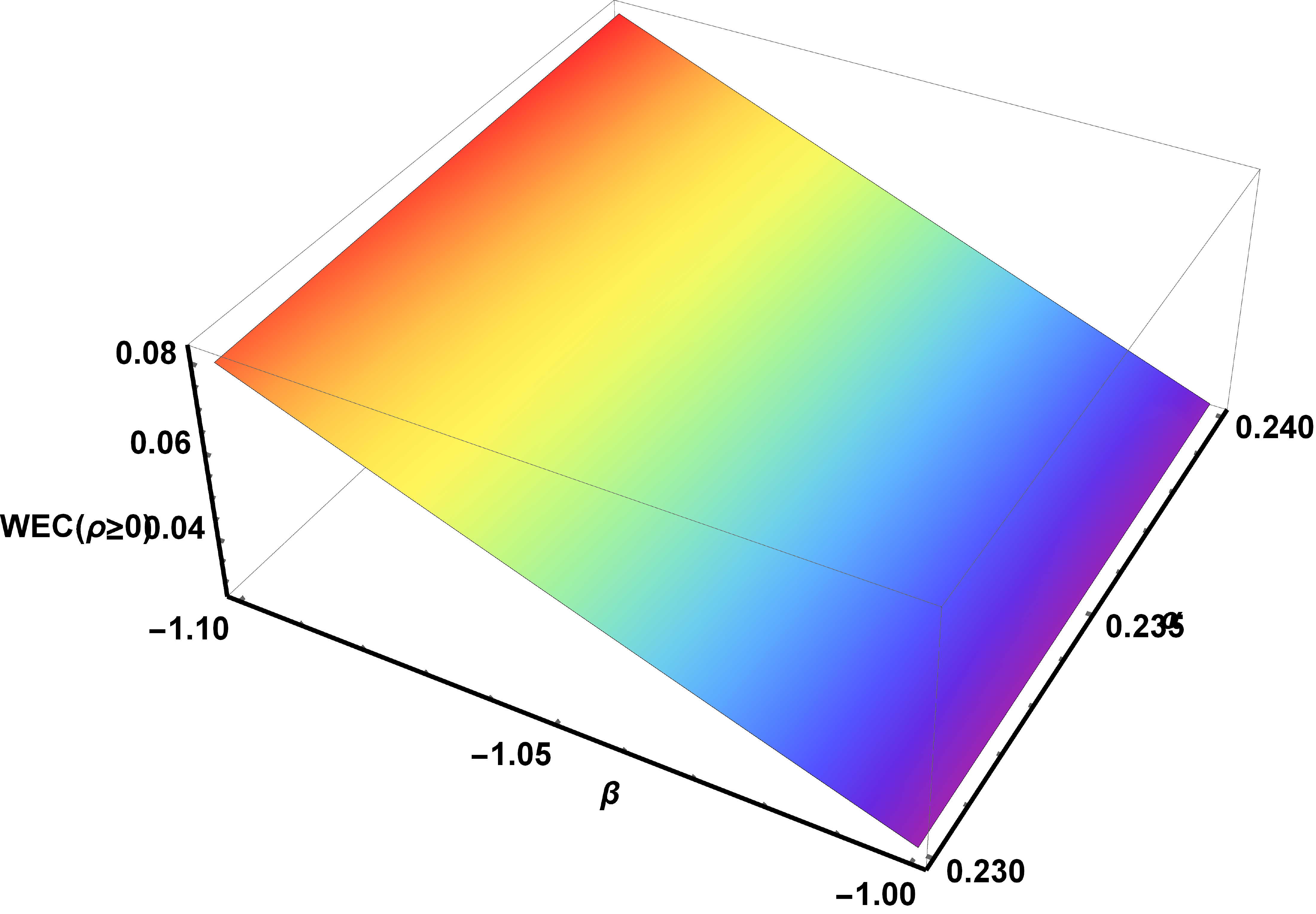}  
  \includegraphics[width=7.5 cm]{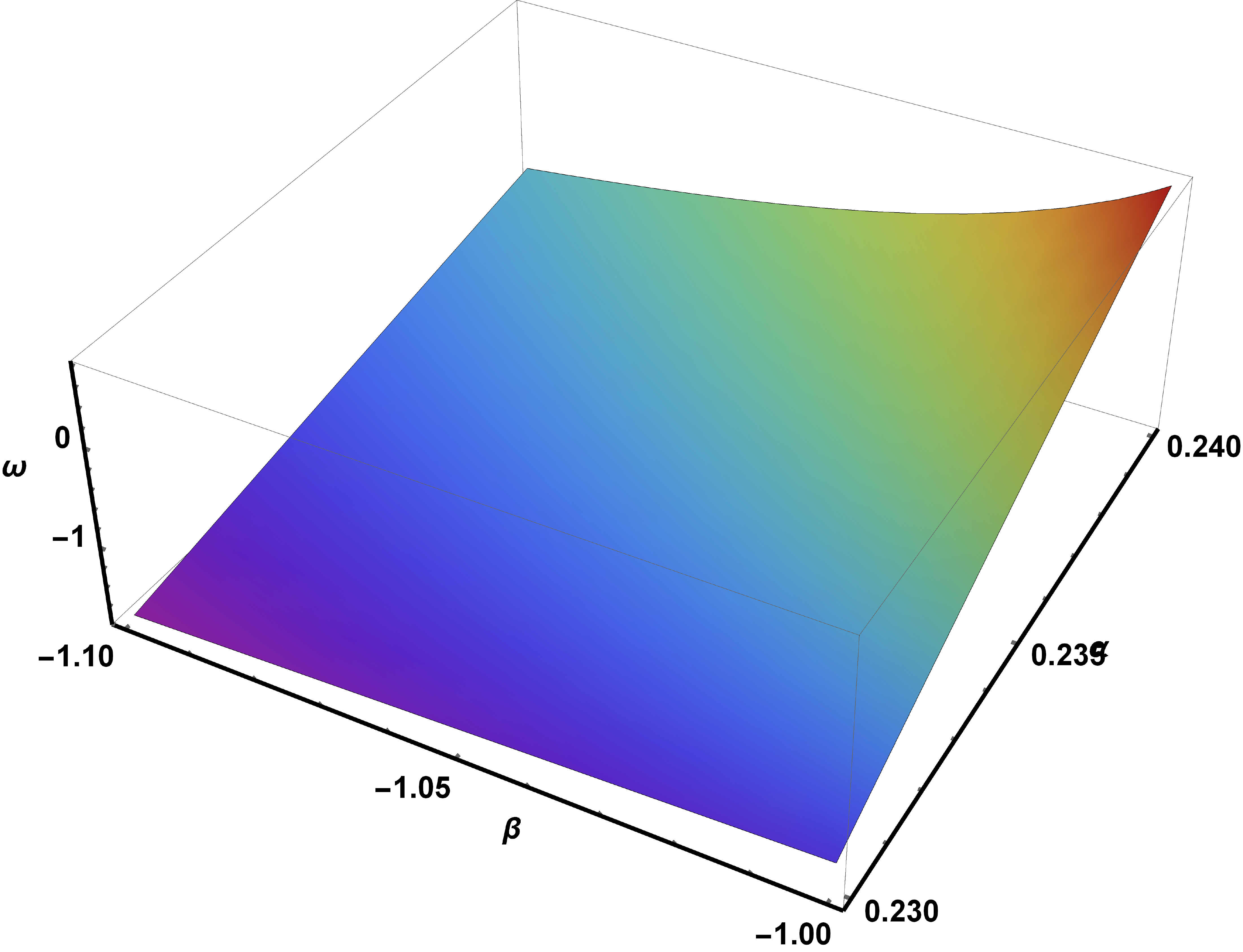}
  \includegraphics[width=7.5 cm]{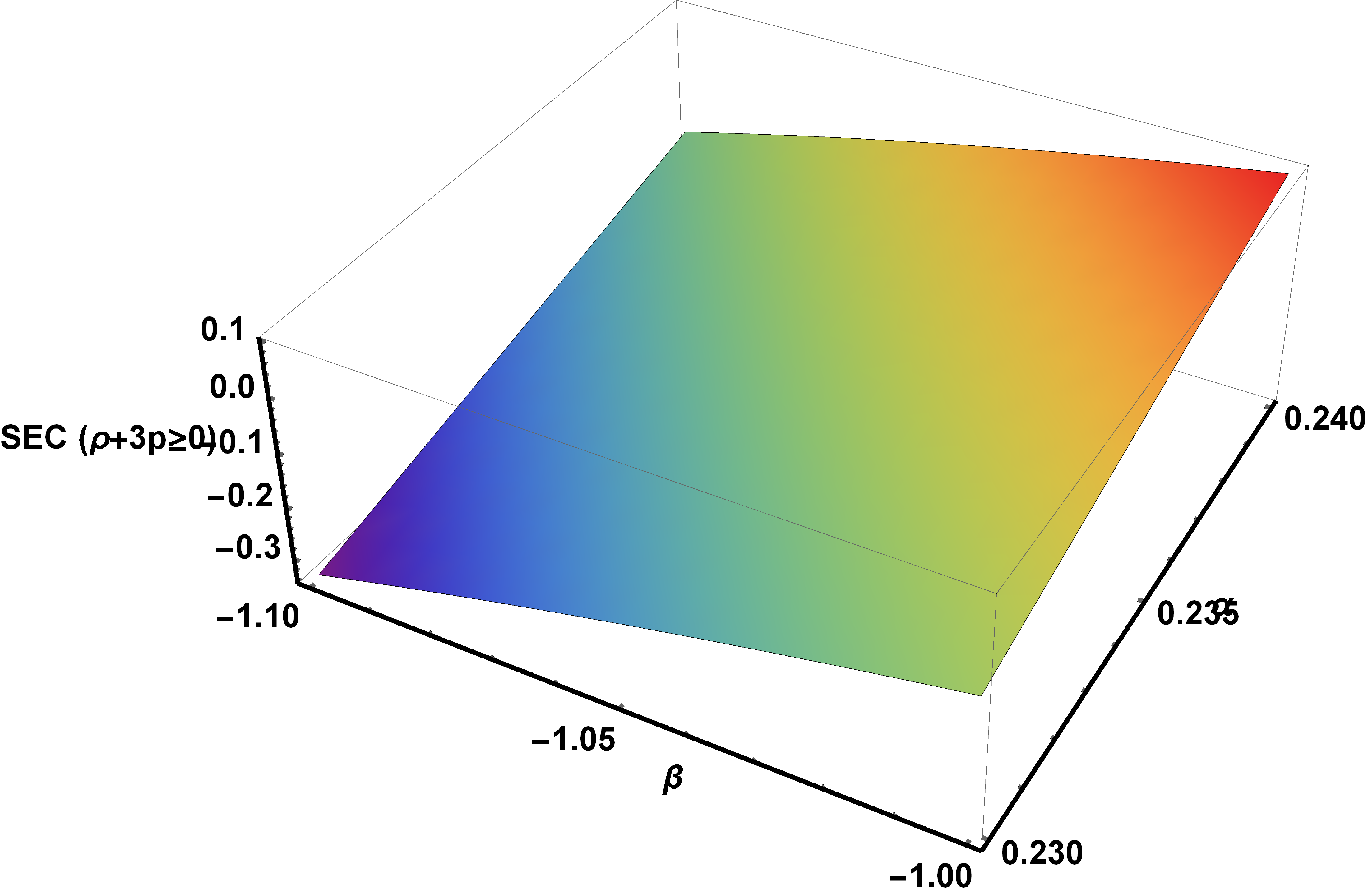}
  \includegraphics[width=7.5 cm]{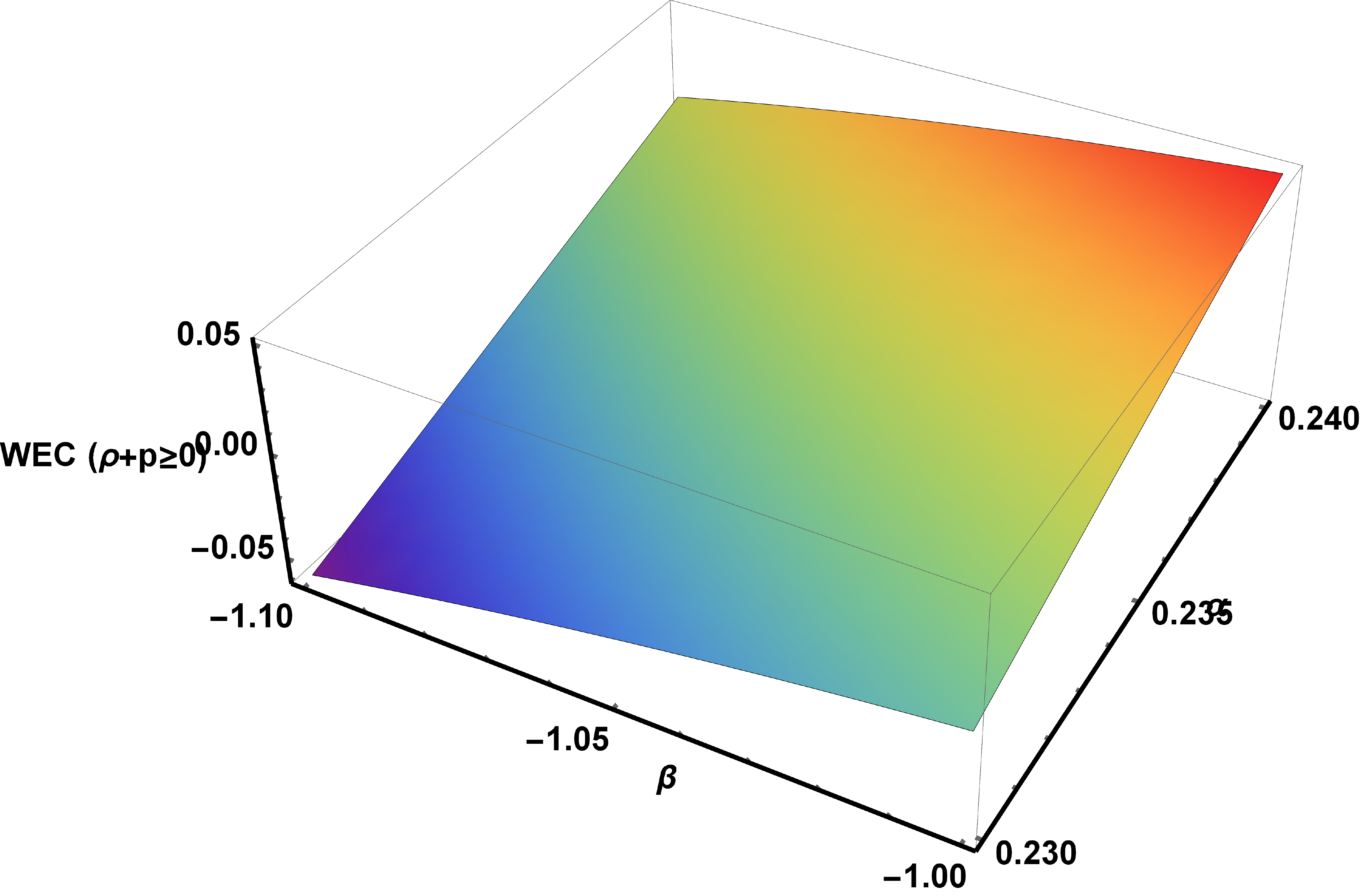}
\caption{Top left panel shows the profile of energy density $\rho_{f(P)}$, top right panel shows the profile of the EoS parameter $\omega_{f(P)}$, bottom left panel shows the profile of the Strong Energy Condition and the bottom right panel shows the profile of the Weak Energy Condition for the second ($f(P) = \alpha \exp (P)$) model. The figures are drawn for $ \beta \left\lbrace -1.1 , -1 \right\rbrace$, and $ \alpha  \left\lbrace 0.23, 0.24 \right\rbrace$. }
\label{fig2}
\end{figure}
 
\section{Conclusions}\label{sec5}
 
Einsteinian cubic gravity (ECG) \cite{p31} is a class of modified gravity theories employing cubic contractions of the Riemann tensor and have undergone rapid and significant development in recent years. $f(P)$ gravity is a novel extension of ECG in which the Ricci scalar in the action is replaced by a function of the curvature invariant $P$ which represents the contractions of the Riemann tensor at the cubic order \cite{p}. This class of modified gravity theory have been very successful in several cosmological domains such as inflation \cite{p35}, spherically symmetric black hole solutions \cite{p32,p33,p34}, and late-time acceleration \cite{p,2p}.\\
The present work is concentrated on bounding some $f(P)$ gravity models to understand their viability in cosmology. Energy conditions are interesting linear relationships between pressure and density and have been extensively employed to derive interesting results in Einstein's gravity, and are also an excellent tool to impose constraints on any cosmological model. To place the bounds, we have ensured that the energy density (Eq. \ref{rho}) must remain positive and the pressure (Eq. \ref{p}) must be negative to assure that the constraints on $f(P)$ gravity models are compatible with the current accelerated expansion of the Universe, and have also investigated to find corners in parameter spaces for a chosen model which permit the EoS parameter (Eq. \ref{om}) to attain a value close to $-1$ to ensure that the bounds are in harmony with the current observational data. \\
In this work, we report bounds for two specific $f(P)$ gravity models where the functional forms of $f(P)$ are represented as \textbf{a)} $f(P) = \alpha \sqrt{P}$, and \textbf{b)} $f(P) = \alpha \exp (P)$, where $\alpha$ is the sole model parameter for these models. We report that for the first model, in order to obtain $\rho_{f(P)}\geq 0$, $p_{f(P)} \leq 0$, and $ \omega_{f(P)}\simeq -1$, the model parameter $\alpha$ must be negative ($\alpha \leq 0$), and $\beta$ must be positive ($\beta \geq 0$), and $\frac{\alpha}{\sqrt{\beta}} \simeq -0.2$, while for the second model, we could not report any conclusive constraints on $\beta
$ and $\alpha$ owing to the complexity of the Eqs. \ref{rho4}, \ref{p4}, and \ref{om4}. Therefore, in order to obtain some qualitative understanding of the parameters $\alpha$ and $\beta$, we proceeded by assuming first $\beta = +1$, and then $\beta = -1$ and checked for which of these two cases, we get $\rho_{f(P)} \geq 0$, and $p_{f(P)} \leq 0$. We found that when $\beta = +1$, to obtain  $\rho_{f(P)}\geq 0$, and $p_{f(P)} \leq 0$, $\alpha$ needs to be positive and negative at the same time which is unphysical and therefore makes it clear-cut that positive values of $\beta$ are not permissible for this model. However, for $\beta=-1$, there breathes suitable values for $\alpha$ which satisfy the aforementioned conditions and also assures that $\omega_{f(P)}\simeq -1$ and reads $\alpha \simeq 0.23$. Furthermore, we report from Figs. \ref{fig1}, and \ref{fig2} that for both the models, the SEC get violated and therefore offered a consistency check to the constraints obtained for both the models.


\end{document}